\documentclass[sigconf,screen]{acmart}

\setcopyright{none}
\acmDOI{}
\acmISBN{}
\acmConference[Tools for Thought Workshop]{CHI '25: Proceedings of the 2025 CHI Conference on Human Factors in Computing Systems}{April 26-May 01, 2025}{Yokohama, Japan}

\begin{document}

\title[Augmenting Expert Cognition in the Age of Generative AI]{Augmenting Expert Cognition in the Age of Generative AI: Insights from Document-Centric Knowledge Work}

\author{Alexa Siu}
\affiliation{
  \institution{Adobe Research}
  \city{San Jose}
  \state{CA}
  \country{USA}
}
\email{asiu@adobe.com}

\author{Raymond Fok}
\affiliation{
  \institution{University of Washington}
  \city{Seattle}
  \state{WA}
  \country{USA}
}
\email{rayfok@cs.washington.edu}

\begin{abstract}
As Generative AI (GenAI) capabilities expand, understanding how to preserve and develop human expertise while leveraging AI's benefits becomes increasingly critical. Through empirical studies in two contexts---survey article authoring in scholarly research and business document sensemaking---we examine how domain expertise shapes patterns of AI delegation and information processing among knowledge workers. Our findings reveal that while experts welcome AI assistance with repetitive information foraging tasks, they prefer to retain control over complex synthesis and interpretation activities that require nuanced domain understanding. 
We identify implications for designing GenAI systems that support expert cognition. These include enabling selective delegation aligned with expertise levels, preserving expert agency over critical analytical tasks, considering varying levels of domain expertise in system design, and supporting verification mechanisms that help users calibrate their reliance while deepening expertise.
We discuss the inherent tension between reducing cognitive load through automation and maintaining the deliberate practice necessary for expertise development. Lastly, we suggest approaches for designing systems that provide metacognitive support, moving beyond simple task automation toward actively supporting expertise development. This work contributes to our understanding of how to design AI systems that augment rather than diminish human expertise in document-centric workflows.
\end{abstract}


\received{20 February 2025}
\received[accepted]{15 March 2025}

\maketitle

\section{Introduction}

As Generative AI (GenAI) capabilities expand, knowledge workers face a pivotal moment in how they interact with information and develop expertise~\cite{lee2025impact, van2023ai, brachman2024knowledge, cambon2023early}. GenAI offers much potential to automate cognitive tasks---from processing large collections of documents to generating comprehensive summary reports and detecting patterns in them~\cite{fok2024marco, liu2024selenite, lo2023semantic}. However, this automation also poses critical questions about maintaining and developing human expertise. 
We examine how domain expertise shapes patterns of AI delegation and information processing through empirical studies of knowledge workers. Our analysis focuses on two contexts that represent different aspects of document-centric work:
(1) survey article authoring and updating in scholarly research, where experts synthesize literature and identify research opportunities~\cite{fok2025livingNarrativeReviews}, and (2) sensemaking over large document collections in business workflows, where experts extract information to support decision-making and reporting~\cite{fok2024marco}.

Prior work suggests that expertise represents more than just accumulated knowledge or skill --- it is fundamentally about how professionals mediate between knowledge and action~\cite{lewandowsky2007knowledge, schubert2013characterizing}. Specifically, experts develop sophisticated mental models and perceptual skills that enable them to maintain flexible cognitive frames throughout their sensemaking process, allowing them to recognize meaningful patterns and adapt their understanding as new information emerges~\cite{schubert2013characterizing, klein2007data}. In contrast, novices are more likely to commit prematurely to initial frames and struggle to incorporate information that doesn't fit their early assumptions~\cite{schubert2013characterizing}. In domains ranging from legal analysis to academic research, knowledge workers must sift through, synthesize, and make decisions based on large collections of documents. This work requires complex cognitive capabilities: developing and adapting frames as new information is acquired, recognizing subtle patterns across documents, and making nuanced judgments based on domain knowledge.

Typically, these cognitive capabilities are developed through deliberate practice and hands-on experience in reading, analyzing, and synthesizing information~\cite{ericsson2004deliberate}. As GenAI makes it possible to automate many aspects of document processing and analysis, we must carefully consider how to preserve and augment opportunities for expert judgment and synthesis. Through our studies on authoring and updating research survey articles and decision-making within business document workflows, we identify key patterns in how expertise shapes AI interaction --- from calibrating trust through AI verification to changes in sensemaking strategies that selectively delegate more information foraging to AI. Our discussion can help suggest ways to design AI systems that support human cognition to enhance productivity, while preserving crucial opportunities for expert analysis, decision-making, and expertise development in document-centric workflows.

\section{Empirical Understanding of GenAI for Document-Centered Knowledge Work}

\subsection{Background}
\subsubsection{Study 1: Sensemaking over large document collections in business workflows}
In the first study, we examined the potential for an interactive, GenAI-enabled system to automate repetitive tasks in knowledge work within business decision-making. The system integrated various features, including lexical and semantic search, extraction, and summarization to facilitate analysis across collections of complex business documents, such as financial reports and legal agreements~\cite{fok2024marco}. 
By structuring information into tables and preserving the provenance of extracted snippets, the system helped reduce cognitive load and tedium, allowing experts to devote more effort to nuanced processes of analyzing and deriving insight from the information identified by GenAI. Results from a controlled usability study found participants completed information-seeking tasks 16\% faster and with less effort using our proposed system when compared to a manual approach without GenAI. One participant reflected on how the system reduced their overall workload, saying \textit{``I definitely feel it saves a lot of my time of going back and forth in the document to search stuff, so it’s really helping me in the kind of mental demand and the amount of work I need to do.''} (P1).

The findings also shed light how GenAI outputs can be inherently imperfect, and often required human verification to calibrate trust and avoid the risks of hallucinations or over-reliance. We observed two common strategies participants used to build confidence when working with GenAI and to recover from errors. First, we observed participants often used different actions in concert for a single query as a means to cross-check retrieved results and surface any inconsistencies. Domain experts described how in workplace contexts where complex information tasks drive high-impact decisions, multiple people would be working together to verify information in different ways, and GenAI could provide an additional redundant layer of review to inform their decision making. Second, we observed when participants had low confidence or encountered any inconsistency, most opted to manually verify the accuracy of AI responses. A key aspect of the system was that it organized results from action queries in a table of results that could be sorted and furthermore allowed users to link each generated response to evidence directly highlighted in each document. This structure allowed users to quickly skim results across several documents and link to specific documents for additional verification when needed. In the controlled usability study, few participants completed the tasks without opening a document and we found no significant differences in task accuracy and self-reported confidence when using our GenAI system compared to the baseline. These observations suggest the system provided adequate mechanisms for users to build trust and reliance on delegated tasks, despite not manually reading and reviewing each document individually.

\subsubsection{Study 2: Survey article authoring and updating in scholarly research}
Our second study investigated how scholars author and update survey articles in computing research~\cite{fok2025livingNarrativeReviews}. Through interviews with survey authors, the study identified three distinct types of updates scholars need to consider: empirical updates involving revision of quantitative data and evidence, structural updates modifying paper organization and taxonomies, and interpretative updates that revise syntheses and narrative framing. While scholars welcomed the potential for AI to assist with routine empirical updates like data extraction and recalculation, they emphasized that the interpretative aspects of surveys---critical for constructing meaningful narratives and maintaining scholarly rigor---remain deeply dependent on expert judgment.

The findings highlighted how understanding emerging research trends requires insider knowledge that goes beyond simple information extraction or summarization. For instance, survey authors described needing deep domain expertise to identify subtle patterns across papers, evaluate the significance of new methodological approaches, and synthesize disparate findings into a coherent narrative that advances the field. Authors expressed particular skepticism about AI's ability to handle these nuanced interpretive tasks, noting that high-quality surveys go beyond an aggregation of existing knowledge to provide critical perspectives that help shape future research directions. This suggests an opportunity for GenAI systems to act as collaborators that handle routine updates while leaving the high-level synthesis and reflective critique to human experts. Survey authors believed that tasks like constructing taxonomies, identifying research gaps, and projecting future directions require not just comprehensive knowledge of published work, but also an understanding of ongoing unpublished research and informal discussions within the research community.

\subsection{Implications to Augment and Support Expert Cognition}

The two studies in diverse domains reveal important patterns in how domain expertise shapes effective human-AI collaboration in knowledge work and in particular when working with documents where provenance is critical. Both studies highlight expertise is deeply contextual and may be challenging to fully automate. For example, survey authors highlighted how understanding emerging research trends requires insider knowledge that goes beyond simple information extraction captured within the artifacts or documents they are working with. Similarly, within business analytics, analysts spend time in information foraging but ultimately their nuanced interpretation is required for making sense of the information, whether it be making a critical business decision or providing a detailed analysis report. We highlight four key implications for designing GenAI systems that support expert cognition:

\paragraph{AI systems should enable selective delegation aligned with expertise levels} In both studies, experts strategically welcomed offloading repetitive, more routine tasks to AI; particularly activities related to information foraging such as document screening, extraction, and structuring. While these foraging activities were necessary for grounding subsequent sensemaking, experts recognized them as tedious and time-consuming. Providing mechanisms to delegate these activities to GenAI can free up cognitive resources to focus more on higher-level analysis and decision making.

\paragraph{AI systems should preserve expert agency over critical analytical tasks}
Experts consistently preferred to retain control over activities that built on the extracted information, such as complex synthesis and interpretation which often required nuanced understanding of values and language semantics. This suggests AI systems should complement rather than replace expert judgment in these areas. For example, survey authors emphasized that once a meaningful narrative or framework has been identified by expertise, the subsequent tasks to fill-in the information are more prone to be delegated. 

\paragraph{AI systems should consider varying levels of domain expertise} While both studies examined knowledge workers with deep domain experience and well-developed analytical capabilities, different user groups may require distinct forms of AI support. These experts demonstrated an ability to adapt, question, and evolve their understanding throughout the sensemaking process, maintaining what prior work calls an open `frame.' However, novices may require different forms of AI support to help them develop similar capabilities. Future work should examine how AI systems can scaffold expertise development while supporting both expert and novice workflows.

\paragraph{AI systems should support verification for calibrating reliance and deepening expertise} The studies show that effective verification of AI outputs requires different strategies depending on the nature of the task. For routine tasks, experts valued quick verification mechanisms such as the ability to trace back to highlighted evidence in source documents.
On the other hand, for analytical tasks involving synthesis or complex interpretations, verification requires deeper engagement with the material. In these cases, experts employed sophisticated strategies, such as using multiple queries in concert to cross-check results and surface inconsistencies, while also relying more on their own judgment to evaluate AI insights. 
Verification serves two important functions: calibrating reliance on AI and deepening understanding of the content. However, as these systems become more capable of connecting and synthesizing larger volumes of information, simple verification methods may be insufficient. For instance, when AI generates abstract insights or identifies complex patterns, checking individual sources may not validate higher-level interpretations. This calls for new verification mechanisms that can enhance AI transparency, while helping users develop appropriate levels of reliance based on task complexity and their own expertise.

\section{Discussion}

There is an inherent tension between reducing cognitive load with GenAI and preserving the hands-on engagement with knowledge needed to develop expertise. Excessive automation risks deskilling, while insufficient support might leave experts unnecessarily spending time on lower-level tasks rather than attending to the complex cognitive activities where their expertise may be more valuable.

Our findings suggest two distinct challenges requiring different forms of metacognitive support~\cite{tankelevitch2024metacognitive}. First, for \textit{experts}, deep domain knowledge can sometimes lead to cognitive entrenchment---where extensive experience may create fixed mental models that limit consideration of alternative approaches or novel solutions. 
Studies suggest experts demonstrate specific strategies to avoid fixation---they may actively reflect on their own thought processes, deliberately seek out alternative perspectives through team collaboration, or systematically analyze problems using structured methods to break free from established mental models~\cite{crilly2015fixation}.
Here, GenAI systems could provide metacognitive scaffolding to help experts reflect on and potentially revise their analytical strategies. For instance, systems could highlight patterns in experts' document analysis approaches, surface potential blind spots that have developed from routine automation, or suggest alternative perspectives that may have been overlooked due to established mental models.

Second, for \textit{novices} or practitioners developing expertise, the framework of deliberate practice suggests that expertise development requires sustained engagement with increasingly challenging tasks, coupled with immediate feedback and opportunities for reflection~\cite{ericsson2004deliberate}. This raises concerns about how automation might limit these crucial learning processes. When working with documents, excessive delegation of information foraging tasks to AI could prevent users from developing deep understanding of the content and hinder their ability to develop effective sensemaking frames through which to interpret the information. To address this, GenAI systems could instead be designed to scaffold expertise development through guided practice rather than purely focus on reducing cognitive load. The fundamental models could be pedagogically steered towards using
effective teaching strategies~\cite{puech2024towards, jurenka2024towards}, and interactive systems could incorporate adaptive interfaces that gradually increase user autonomy as competence grows, while providing explicit prompts for reflection on decision-making processes, reflecting active learning successes in student education~\cite{freeman2014active, merrill1992effective}.
By offering contextual explanations, such systems can help users understand not just what information is relevant, but \textit{why}, and enable structured comparison of user analyses with AI-generated alternatives to promote deeper understanding. Critical to this approach is the progressive introduction of automation features (or delegation scope) with respect to demonstrated skill development.

The goal is to move beyond simple task automation toward systems that actively support metacognitive development, helping both experts and novices develop more sophisticated mental models and analytical capabilities. This suggests a shift in how we conceptualize GenAI assistance: \textit{from purely reducing cognitive load to strategically engaging users in ways that promote expertise development} while still leveraging automation's efficiency benefits. By carefully designing these metacognitive supports, we can help ensure that AI augmentation enhances rather than diminishes the development of human expertise.

Future research could further study how different levels of automation affect expertise retention and transfer. For example, what types of GenAI integrations best preserve the depth of expert knowledge while accelerating present workflow efficiency. 
Does GenAI-assisted synthesis change how authors internalize and recall knowledge compared to manually reading, organizing, and interpreting the same information? These questions remain at the forefront as we continue to design GenAI systems that strive to protect and augment humans' complex cognition over time.

\bibliographystyle{ACM-Reference-Format}
\bibliography{main}


\end{document}